\documentclass[preprint2]{aastex}

\shorttitle{Comment on arXiv:0705.1508}
\shortauthors{Alister W.\ Graham}

\begin{document}

\title{Comment on arXiv:0705.1508}

\vspace{-1.1cm}

\author{Alister W.\ Graham\altaffilmark{1}}
\affil{Centre for Astrophysics and Supercomputing, Swinburne University
of Technology, Hawthorn, Victoria 3122, Australia.}
\altaffiltext{1}{Corresponding Author: AGraham@astro.swin.edu.au}

\vspace{-1.1cm}

\begin{abstract}
  
  A simple regression analysis designed for predicting the supermassive black
  hole mass from the effective radius and mean effective surface brightness of
  the host bulge has been performed using the data from Barway \& Kembhavi
  (arXiv:0705.1508).  The scatter in the $\log M_{\rm bh}$ direction is found
  to be 0.32 dex, 
  comparable with values
  obtained using a single predictor quantity such as luminosity, velocity
  dispersion or S\'ersic index.

\end{abstract}


\section{Research Note}

\begin{figure}
\includegraphics[angle=270,scale=0.38]{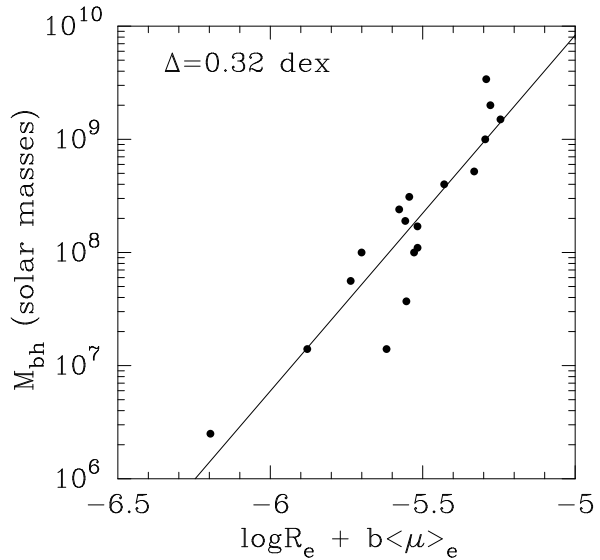}
\caption{
Regression analysis of $\log M_{\rm bh}$ against $\log R_{\rm e}$ and $\langle
\mu \rangle _{\rm e}$.  The quantity $b$ is found to equal $-0.29$
(equation~\ref{Eq_18}). 
}
\label{Figer}
\end{figure}

Barway \& Kembhavi (2007) have made the interesting claim that a combination
of two photometric parameters, namely the effective radius $R_{\rm e}$ and 
the mean effective
surface brightness $\langle \mu \rangle _{\rm e}$, 
can be used to predict supermassive black hole masses with
a greater degree of accuracy than single quantities such as velocity
dispersion (Ferrarese \& Merritt 2000; Gebhardt et al.\ 2000), 
luminosity (Graham 2007 and references therein) 
or major-axis S\'ersic index (Graham \& Driver 2007).  These latter 
relations have a total scatter in the $\log M_{\rm bh}$ direction of
0.31--0.34 dex. 

Here we check Barway \& Kembhavi's (2007) claim that the total scatter 
in the $\log M_{\rm bh}$ direction, when using $\langle \mu \rangle _{\rm e}$ 
and $\log R_{\rm e}$ as predictor quantities, 
is 0.25 dex (and 0.19 dex when excluding the outlier NGC~4742). 

A simple ordinary least squares regression analysis of 
$(Y$$\mid$$X)$ has been performed, in which 
$Y=\log M_{\rm bh}$ and 
$X=\log R_{\rm e} + b\langle \mu \rangle _{\rm e}$.  
The data for $\log M_{\rm bh}$, $\log R_{\rm e}$ and $\langle \mu \rangle _{\rm e}$
have been taken from Table~1 in Barway \& Kembhavi (2007).   
In addition to solving for the parameter $b$ in the above inset equation, 
we also solve for $A$ and $B$ to give the expression 
\begin{equation}
Y = A + BX.
\end{equation}
Given the absence of reported errors on the quantities $\log R_{\rm e}$ and $\langle
\mu \rangle _{\rm e}$ in Barway \& Kembhavi (2007), no attempt has been made
to include such measurement errors
in the regression, and subsequently no attempt to quantify the intrinsic
scatter has been made. 

The optimal solution (using all 18 data points) is 
\begin{equation}
\log M_{\rm bh} = 25.65 + 3.15\log R_{\rm e} - 0.90\langle \mu \rangle _{\rm e}
\label{Eq_18}
\end{equation}
and is shown in Figure~\ref{Figer}.  The scatter in the $\log M_{\rm bh}$
direction is 0.32 dex.  

Upon the exclusion of NGC~4742, the optimal relation is
\begin{equation}
\log M_{\rm bh} = 27.88 + 3.23\log R_{\rm e} - 1.01\langle \mu \rangle _{\rm e}, 
\end{equation}
and the scatter is reduced to 0.25 dex --- which is the same level of scatter in 
the $M_{\rm bh}$--$n$ relation upon the removal of two outliers (Graham \&
Driver 2007).  This level of scatter is also equal to the value reported 
by Marconi \& Hunt (2003) who also used a combination of two 
parameters ($R_{\rm e}$ and $\sigma$) to predict $M_{\rm bh}$. 

The low value of 0.19 dex reported by Barway \& Kembhavi appears to have
arisen by dividing the scatter in the $\log R_{\rm e}$ direction (0.061)
by the coefficient in front of the $\log M_{\rm bh}$ term in their equation~3
(which is their fitted plane).  This overlooks the three-dimensional nature 
of the plane and consequently results in the over-estimation of the plane's
ability to predict black hole masses.  An easy check is to compute
the offset between the black hole masses listed in Table~1 of 
Barway \& Kembhavi and the values predicted from their plane (their 
equation~3), which 
can be re-written as
\begin{equation}
\log M_{\rm bh} = 27.16 + 3.13\log R_{\rm e} - 0.97\langle \mu \rangle _{\rm e}. 
\end{equation}
Doing so (and excluding NGC~4742) the total rms scatter
in the $\log M_{\rm bh}$ direction is 0.27 dex (not 0.19 dex, and 
greater than the value of 0.25 dex obtained using the optimal 
plane constructed here).

In passing it is noted that three of the galaxies used by
Barway \& Kembhavi are known to 
be disc galaxies, or at least are not regular elliptical galaxies.
M32 may likely be a stripped S0 galaxy (e.g.\ Graham 2002), while 
NGC 2778 is a disc galaxy (e.g. Rix, Carollo \& Freeman 1999), as is 
NGC 4564 (Trujillo et al.\ 2004; 
see also figure~6 in Graham \& Driver 2007).  Consequently, the 
effective radii and mean surface brightnesses which have been
used for these galaxies do not pertain to the bulge.  
Curiously, excluding these three galaxies (while including NGC~4742), 
the rms scatter is 0.33 dex.

%

\label{lastpage}
\end{document}